\documentclass[aps, prl, twocolumn, showpacs, superscriptaddress, floatfix]{revtex4}

\usepackage{graphicx}
\usepackage{dcolumn}
\usepackage{bm}
\usepackage{times}
\usepackage{amsmath}

\begin{document}

\title{Magnetic ordering in the frustrated $J_1$-$J_2$ Ising chain candidate BaNd$_2$O$_4$}

\author{A.A. Aczel}
\altaffiliation{author to whom correspondences should be addressed: E-mail:[aczelaa@ornl.gov]}
\affiliation{Quantum Condensed Matter Division, Oak Ridge National Laboratory, Oak Ridge, TN 37831, USA}
\author{L. Li}
\affiliation{Department of Materials Science and Engineering, University of Tennessee, Knoxville, TN 37996, USA}
\author{V.O. Garlea}
\affiliation{Quantum Condensed Matter Division, Oak Ridge National Laboratory, Oak Ridge, TN 37831, USA}
\author{J.-Q. Yan}
\affiliation{Department of Materials Science and Engineering, University of Tennessee, Knoxville, TN 37996, USA}
\affiliation{Materials Science and Technology Division, Oak Ridge National Laboratory, Oak Ridge, TN 37831, USA}
\author{F. Weickert}
\affiliation{MPA-CMMS, Los Alamos National Laboratory, Los Alamos, NM 87545, USA}
\author{M. Jaime}
\affiliation{MPA-CMMS, Los Alamos National Laboratory, Los Alamos, NM 87545, USA}
\author{B. Maiorov}
\affiliation{MPA-CMMS, Los Alamos National Laboratory, Los Alamos, NM 87545, USA}
\author{R. Movshovich}
\affiliation{MPA-CMMS, Los Alamos National Laboratory, Los Alamos, NM 87545, USA}
\author{L. Civale}
\affiliation{MPA-CMMS, Los Alamos National Laboratory, Los Alamos, NM 87545, USA}
\author{V. Keppens}
\affiliation{Department of Materials Science and Engineering, University of Tennessee, Knoxville, TN 37996, USA}
\author{D. Mandrus}
\affiliation{Department of Materials Science and Engineering, University of Tennessee, Knoxville, TN 37996, USA}
\affiliation{Materials Science and Technology Division, Oak Ridge National Laboratory, Oak Ridge, TN 37831, USA}
\affiliation{Department of Physics and Astronomy, University of Tennessee, Knoxville, TN 37996, USA}

\date{\today}

\begin{abstract}
The AR$_2$O$_4$ family (R = rare earth) have recently been attracting interest as a new series of frustrated magnets, with the magnetic R atoms forming zigzag chains running along the $c$-axis. We have investigated polycrystalline BaNd$_2$O$_4$ with a combination of magnetization, heat capacity, and neutron powder diffraction (NPD) measurements. Magnetic Bragg peaks are observed below $T_N$~$=$~1.7~K, and they can be indexed with a propagation vector of $\vec{k}$~$=$~(0 1/2 1/2). The signal from magnetic diffraction is well described by long-range ordering from only one of the two types of Nd zigzag chains, with collinear up-up-down-down intrachain spin configurations. Furthermore, low temperature magnetization and heat capacity measurements reveal two field-induced spin transitions at 2.5~T and 4 T for $T$~$=$~0.46~K. The high field phase is paramagnetic, while the intermediate field state may arise from a spin transition of the long-range ordered Nd chains, resulting in an up-up-down intrachain spin configuration. The proposed intermediate field state is consistent with the magnetic structure determined in zero field for these chains by NPD, as both phases are predicted for the classical Ising chain model with nearest neighbor and next nearest neighbor interactions.
\end{abstract}


\maketitle

\renewcommand{\topfraction}{0.85}
\renewcommand{\textfraction}{0.1}
\renewcommand{\floatpagefraction}{0.7}

\section{\label{sec:level1}I. Introduction}

Geometric frustration is a term used to describe magnetic systems where all of the microscopic exchange interactions cannot be satisfied simultaneously, and as a consequence this class of materials often exhibits exotic magnetic ground states and complex magnetic behavior. Some common architectures characterized by frustration include pyrochlore\cite{10_gardner}, Kagome\cite{07_helton}, and face-centered cubic magnetic sublattices\cite{10_aharen, 10_aharen_2}. Due to the magnetic frustration inherent in these materials, non-collinear magnetic ground states often form that can lead to multiferroic behavior\cite{07_cheong}. In other cases, conventional long-range magnetic order is replaced by short-range alternatives, including spin glasses\cite{14_silverstein}, spin liquids\cite{99_gardner, 00_gingras}, and spin ices\cite{97_harris, 99_ramirez, 00_denhertog}. 

Recently, the family of materials AR$_2$O$_4$ (A~$=$~Ba, Sr; R~$=$~rare earth) have been attracting interest as a new series of frustrated magnets. These systems consist of two crystallographically-inequivalent rare earth sites with very different, distorted octahedral oxygen environments. Each site forms zigzag chains (or ladders) of magnetic R atoms running along the $c$-axis. Bulk probes, including magnetic susceptibility and heat capacity, have shown that most members of the family have dominant antiferromagnetic (AFM) exchange interactions and relatively large frustration indices\cite{05_karunadasa, 06_doi}. Geometric frustration can arise in this structure type if the next nearest neighbor (NNN) $J_2$ intrachain exchange interactions (ladder legs) are AFM and the nearest neighbor (NN) $J_1$ intrachain couplings (ladder rungs) are of comparable strength. 

The phase diagram for both the magnetic $S$~$=$~1/2\cite{89_igarashi} and classical\cite{07_heidrich} Ising-like, AFM $J_1$-$J_2$ chain models have been determined theoretically. In both cases, a simple up-down-up-down (N\'{e}el) phase and an up-up-down-down (UUDD) phase have been predicted depending on the relative values of $J_1$ and $J_2$. For $S$~$=$~1/2 systems in zero field, the two phases are separated by a quantum critical point at $J_2$~$=$~$J_1$/2. An additional canted phase, with finite magnetization, is found between the N\'{e}el and UUDD states in the zero field phase diagram for classical systems. In both the quantum $S$~$=$~1/2 and classical spin limits, the calculations predict a field-induced up-up-down (UUD) phase over a particular $J_2$/$J_1$ range for an applied magnetic field along the ordered spin direction. The UUD spin structure leads to a 1/3 magnetization plateau. Despite the extensive theoretical work on the $J_1$-$J_2$ Ising chain model, there are not many examples that can be used to test its predictions, especially in the classical limit. Some members of the AR$_2$O$_4$ family, in particular SrDy$_2$O$_4$ and SrHo$_2$O$_4$, have recently been proposed as model Ising $J_1$-$J_2$ chain systems\cite{14_poole}. There are also several other classical Ising chain candidates in this family with large total angular momenta $J$ associated with the magnetic R atoms.   

Several recent studies have been performed on SrR$_2$O$_4$, and both the N\'{e}el and UUDD ordering predicted by the $J_1$-$J_2$ Ising chain model have been observed in these classical magnetic spin systems. However, there are several aspects of the magnetism not explained by the original theories. For example, SrEr$_2$O$_4$ exhibits long-range N\'{e}el order\cite{08_petrenko}, characteristic of a small $J_2$, coexisting with diffuse magnetic scattering\cite{11_hayes}. Monte Carlo simulations reveal that the pattern from the latter is best described by dominant $J_2$ AFM exchange interactions, and this finding is consistent with quasi-1D UUDD ordering. One way the different scattering patterns can coexist is if the two types of zigzag chains have inequivalent $J_1$ and $J_2$ interactions and therefore host different magnetic ground states. Support for this interpretation comes from Rietveld refinement of the neutron powder diffraction (NPD) data, as the best refinements arise when the long-range order is only associated with one type of Er zigzag chain\cite{08_petrenko}. 

A wide range of magnetic behavior is found across the SrR$_2$O$_4$ series. SrHo$_2$O$_4$ shows very similar magnetism to the Er analogue with coexisting N\'{e}el and UUDD ordered states, although the correlation length of these states along different crystallographic directions is still under debate\cite{11_young, 13_young, 14_wen}. As in the Er case, the two different structures can likely be associated with the two types of Ho zigzag chains. In contrast, NPD shows that SrYb$_2$O$_4$ is characterized by long-range N\'{e}el order on both Yb sites, although the ordered Yb moment is drastically suppressed on one of the two sites\cite{12_quintero}. Finally, SrDy$_2$O$_4$ shows evidence for quasi-1D, UUDD magnetic correlations in zero field down to 50~mK\cite{14_poole}, while SrTm$_2$O$_4$ shows no signs of magnetic ordering down to 65~mK\cite{14_li}. The low ($T$~$<$~1~K) or non-existent magnetic ordering temperatures, coupled with the variety of observed behavior, suggest that some combination of competing exchange interactions, single ion anisotropy, magnetic dipole interactions, and low dimensionality play important roles in shaping the magnetic phase diagrams of these materials. 

In this work, we expand the earlier studies of these zigzag chain systems to BaNd$_2$O$_4$ ($J$~$=$~9/2) through magnetization, heat capacity, and neutron diffraction measurements on polycrystalline samples. This material has a Curie-Weiss temperature of -24~K, while the measurements presented in this work indicate a magnetic transition of only $T_N$~$=$~1.7 K, indicative of a high magnetic frustration index f~$\sim$~12. Neutron diffraction data show evidence for a long-range antiferromagnetic ground state that arises from only one of the two Nd sites, characterized by a propagation vector of $\vec{k}$~$=$~(0 1/2 1/2) and the spins lying in the $ab$-plane. Furthermore, low temperature magnetization and heat capacity measurements as a function of applied field reveal an intermediate field-induced ordered state, and also indicate that any magnetic order is completely suppressed in an applied field of about 4 T as $T \rightarrow 0$.     

\section{\label{sec:level2}II. Methods}

Single phase polycrystalline BaNd$_2$O$_4$ samples were synthesized by a standard solid-state reaction method from high-purity starting materials of BaCO$_3$ and Nd$_2$O$_3$. About 15\% extra BaCO$_3$ was added to compensate for the evaporation of barium. The starting materials were mixed and ground in an agate mortar and then pressed into pellets. The pellets were sintered in a flowing Ar atmosphere at 1150$^\circ$C for 8 hours. The resultant pellets were then reground and pressed into new pellets for more sintering in the Ar atmosphere at 1300$^\circ$C for 10 hours. The final product was confirmed to be single phase by laboratory x-ray powder diffraction. 

Neutron powder diffraction (NPD) was performed with 5~g of polycrystalline BaNd$_2$O$_4$ at Oak Ridge National Laboratory using the HB-2A powder diffractometer of the High Flux Isotope Reactor. The sample was sealed in an Al sample can with He exchange gas, and a He-3 sample environment was used for the experiment with a base temperature of 300~mK. All measurements were conducted with a neutron wavelength of 2.41~\AA~and a collimation of 12$'$-open-6$'$. The NPD data was analyzed using the Rietveld refinement program FullProf\cite{93_rodriguez} and the representational analysis software SARAh\cite{00_wills}.

The specific heat was measured with a home-built probe based on the adiabatic heat-pulse technique in a 3He/4He dilution refrigerator from Oxford Instruments. The experiment was carried out in a 14 T superconducting magnet. Special care was taken to ensure the proper thermalization of the cold-pressed polycrystalline sample during the low temperature measurements.
 
Magnetization measurements in fields up to 7 T were performed in a Quantum Design MPMS SQUID magnetometer. Data from $\sim$ 1.7 K to 100 K was taken using the standard $^4$He setup. Data between 0.44 K and 2 K was collected with an iQuantum $^3$He insert that fits inside the MPMS sample space and utilizes the same SQUID detection system and procedure. Excellent agreement between the $^3$He and $^4$He data was observed in the temperature range of overlap.   

\section{\label{sec:level3}III. Discussion and Analysis}

\begin{figure}
\centering
\scalebox{0.28}{\includegraphics{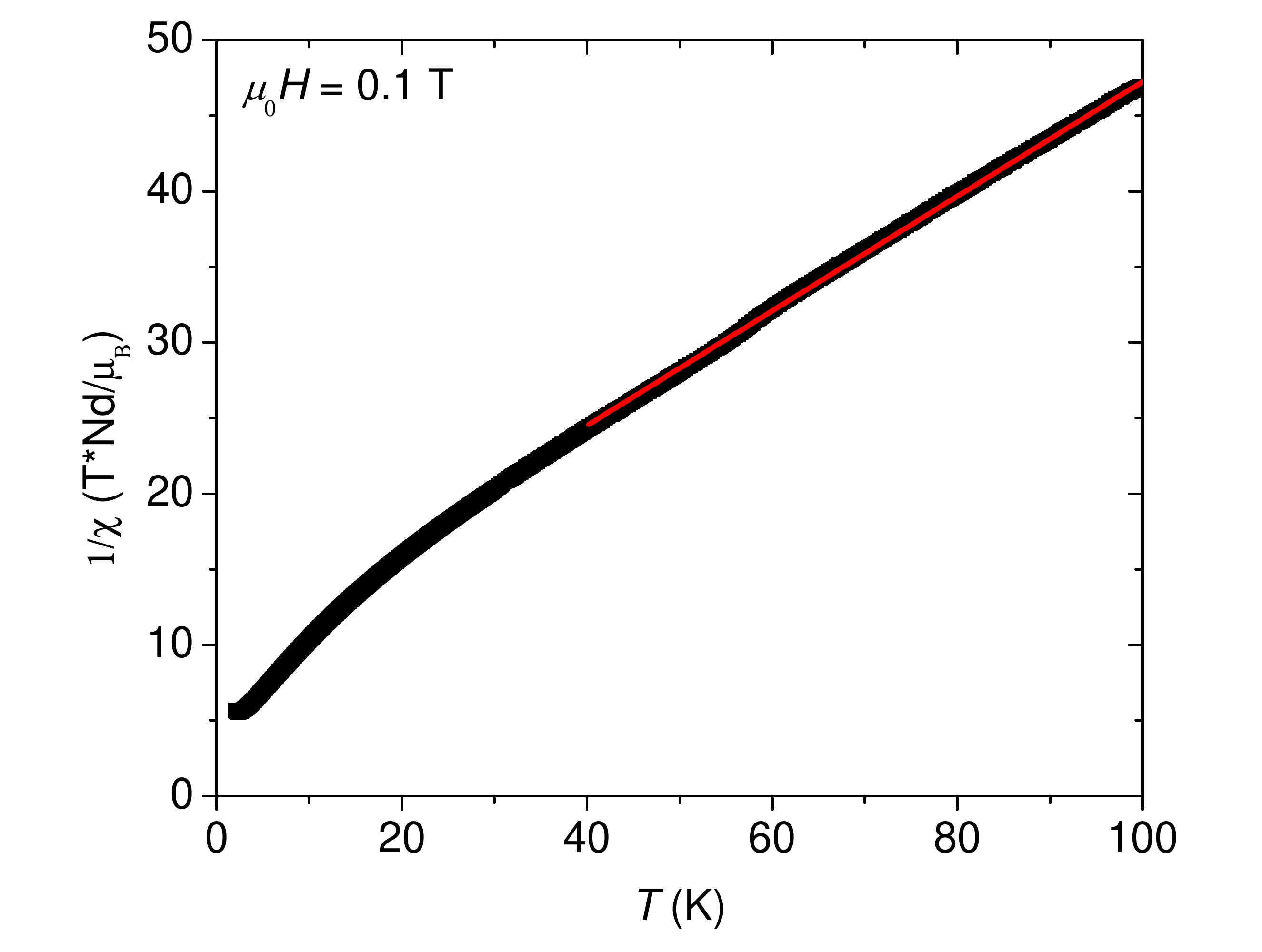}}
\caption{\label{characterization} Temperature dependence of the inverse magnetic susceptibility for BaNd$_2$O$_4$ in an applied field of 0.1~T. The solid red line is a fit to a Curie-Weiss law in the temperature range 40 - 100 K. The broad valley around $T$~$=$~2.2~K likely arises from the onset of an antiferromagnetic transition.}
\end{figure}

Magnetic susceptibility measurements of polycrystalline BaNd$_2$O$_4$ in an applied field $\mu_0 H$~$=$~0.1~T are presented in Fig.~\ref{characterization} (plotted as 1/$\chi$ vs. $T$) and the results are in good agreement with previous work\cite{06_doi}. There was no observable difference for data collected under field-cooled and zero field-cooled conditions, and a broad valley is apparent in 1/$\chi$ around 2.2~K, likely corresponding to the onset of an antiferromagnetic transition. The high temperature data is well-described by a Curie-Weiss law, with the value of the Weiss temperature $\theta_{CW}$ strongly dependent on the chosen fitting range, as first discussed in Ref.~\cite{06_doi}. Fitting the data to the Curie-Weiss law between 40 and 100~K yields $\theta_{CW}$~$=$~-24~K and an effective moment $\mu_{eff}$~$=$~3.40(1)~$\mu_B$. The effective moment is close to the expected value of 3.62~$\mu_B$ for Nd$^{3+}$.

\begin{figure}
\centering
\scalebox{0.43}{\includegraphics{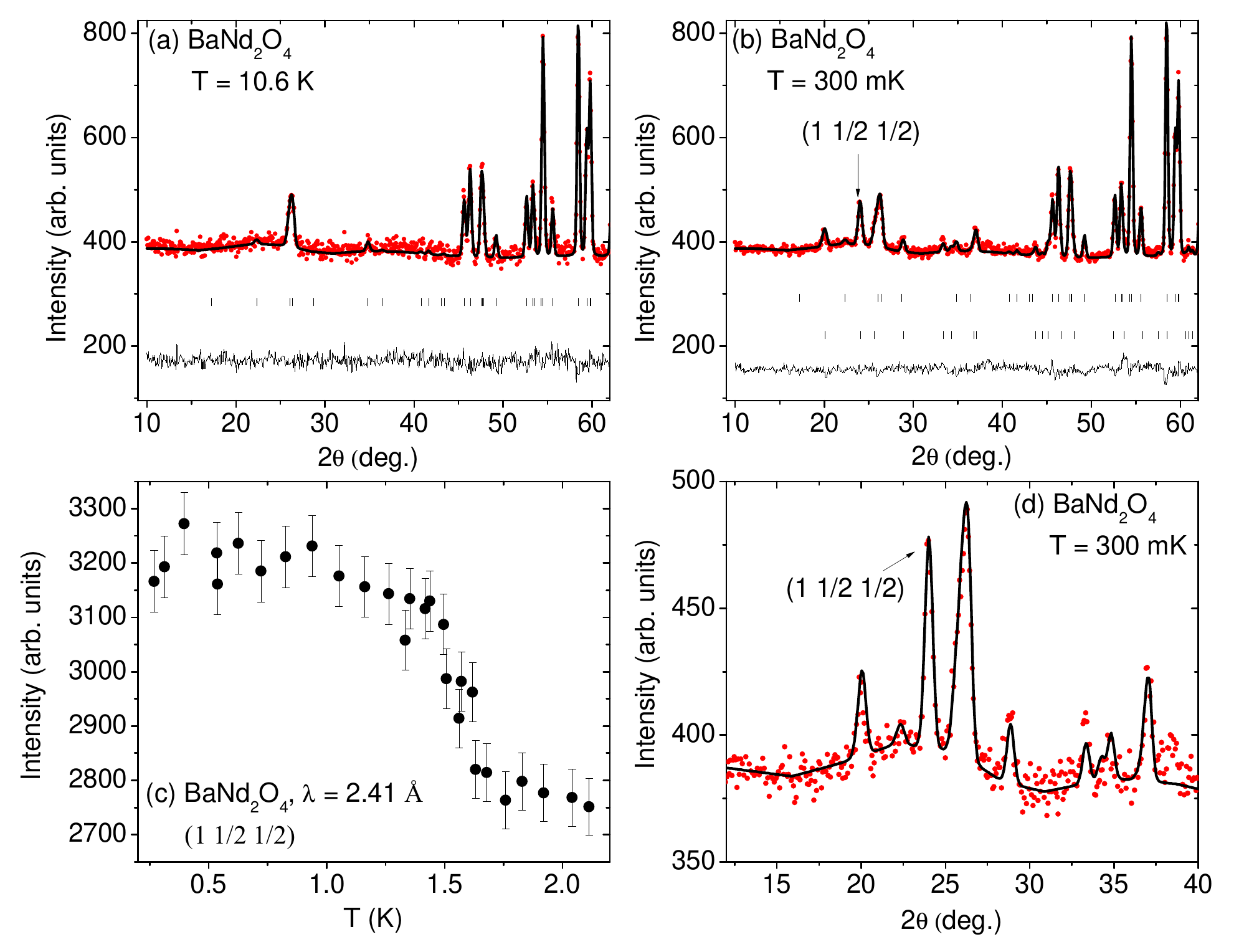}}
\caption{\label{neutron_diffraction} (a), (b) HB-2A neutron diffraction data with $\lambda$~$=$~2.41~\AA~both above and below $T_N$ for polycrystalline BaNd$_2$O$_4$. The solid lines are the fits generated from Rietveld refinements using the space group {\it Pnam}. One of the four canadidate magnetic structures, described in the text, is also incorporated in the refinement shown at 300~mK. (c) Intensity of the (1 1/2 1/2) Bragg peak plotted vs. temperature. (d) An enlarged version of the plot in (b), showing evidence for diffuse scattering centered around $\vec{Q}$~$=$~(1 1/2 1/2).}
\end{figure}

Neutron diffraction data from HB-2A with a neutron wavelength of 2.41~\AA~is depicted in Fig.~\ref{neutron_diffraction} at temperatures both above and below $T_N$. Successful Rietveld refinements were performed with the room temperature space group {\it Pnam}, indicating that there are no structural phase transitions down to 300~mK. The lattice constants at 300~mK refined as $a$~$=$~10.573(1)~\AA, $b$~$=$~12.432(1)~\AA, and $c$~$=$~3.601(1)~\AA. Several new Bragg peaks appeared below $T_N$, as shown in Fig.~\ref{neutron_diffraction}(b), and these could all be indexed with the propagation vector $\vec{k}$~$=$~(0 1/2 1/2). The temperature dependence of the most intense magnetic peak (1 1/2 1/2) is plotted in Fig.~\ref{neutron_diffraction}(c), with the onset clearly observed around $T_N$~$=$~1.7~K, in good agreement with previous work\cite{06_doi}. Diffuse scattering centered about $\vec{Q}$~$=$~(1 1/2 1/2) is also observed in the data, as shown most clearly in Fig.~\ref{neutron_diffraction}(d).

Representational analysis allowed the possible magnetic structures to be constrained on the basis of the crystal symmetry. There are two possible irreducible representations $\Gamma_1$ and $\Gamma_2$ (Kovalev's notation\cite{kovalev}), each with six basis vectors, and therefore there are several possible candidates for the magnetic structure of BaNd$_2$O$_4$. However, when refinements were attempted with several different models, the component of the Nd magnetic moments along $\hat{c}$ was consistently found to be negligible ($<$~0.2~$\mu_B$). This parameter was assumed to be zero in subsequent refinements as a result. 

Recent work on the isostructural SrR$_2$O$_4$ family has emphasized the important role that the different crystal field environments of the two R sites play in governing the magnetic properties of these materials\cite{14_poole, 14_wen}. More specifically, neutron diffraction has generally found that the inequivalent zigzag chains host different types of magnetic order or they are characterized by different ordered moment sizes. In some cases, only one of the two R sites was found to exhibit long-range order\cite{08_petrenko}. For these reasons, refinements were attempted with equal moments on the two sites and also with the moment on one R site constrained to be zero.

\begin{figure*}
\centering
\scalebox{0.77}{\includegraphics{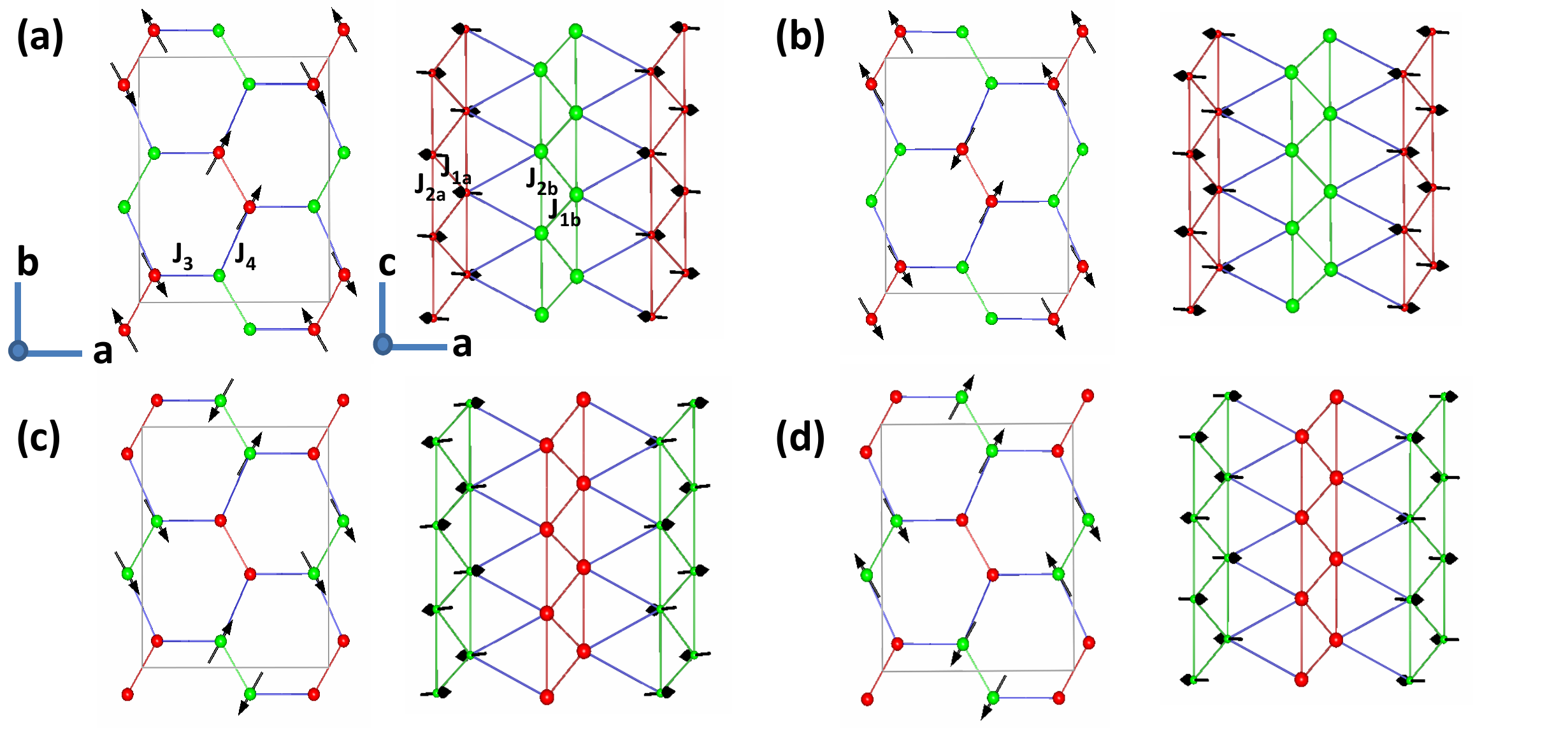}}
\caption{\label{mag_structure} (a)-(d) Four candidate magnetic structures for BaNd$_2$O$_4$, with schematics of the $ab$ and $ac$-planes shown separately. The two crystallographically-inequivalent Nd sites are indicated by the atoms of different colors (red and green), while the two types of zigzag chains are constructed from bonds of different colors (red and green). The series of relevant exchange interactions for the system is indicated in (a).}
\end{figure*}

The best statistical refinements correspond to the scenario where only one Nd site participates in the long-range ordering. There are eight magnetic configurations that yield nearly identical fits of the data. Four of the models consist of collinear UUDD intrachain spin arrangements, which are expected for classical Ising $J_1$-$J_2$ chains in the large AFM $J_2$ limit\cite{07_heidrich}. We note that the related systems SrHo$_2$O$_4$ and SrDy$_2$O$_4$ have already been described as Ising $J_1$-$J_2$ chain systems\cite{14_poole}. For these reasons, the four intrachain collinear structures represent the most likely ordering configurations for BaNd$_2$O$_4$. 

The other four possible models consist of non-collinear intrachain spin textures, with NNN spins along the chains AFM-coupled. These spin structures correspond to helices running along the chain direction for the special case when the NN spins are 90$^\circ$ out of phase (i.e. when four consecutive spins make up a complete period for the helix). While helical order is the expected ground state for an AFM Heisenberg $J_1$-$J_2$ model when $J_2$~$>$~0.25$J_1$\cite{79_deraedt, 88_harada, 07_heidrich}, these non-collinear models only yield equivalent statistical refinements to the UUDD models when the phases between consecutive pairs of NN spins are not constrained and refine to 63.6$^\circ$ and 116.4$^\circ$. The resulting structures are unlikely from a physical standpoint and therefore are discarded. 

NPD cannot determine a unique magnetic structure for BaNd$_2$O$_4$ because the refinements are insensitive to two different features of the system. Firstly, the R site that corresponds to the long-range order cannot be distinguished by this technique. This issue was first discussed for SrEr$_2$O$_4$\cite{08_petrenko}. The  situation for the present case is very similar to the Er system, as only one Er site was found to show long-range order. Secondly, the relative phase along $\hat{c}$ of the UUDD ordered zigzag chains centered around $y$~$=$~0 and $y$~$=$~0.5 remains unknown. Equivalent refinements are obtained if the $J_1$ bonds within the chemical unit cell (i.e. connecting Nd atoms with $z$~$=$~0.25 and $z$~$=$~0.75) are FM and AFM for the $y$~$=$~0 and $y$~$=$~0.5 chains respectively or vice versa. Single crystal polarized neutron diffraction or point charge calculations of the crystal field levels may be able to help determine the type of long-range magnetic order in BaNd$_2$O$_4$ unambiguously.

The four intrachain collinear structures are presented in Fig.~\ref{mag_structure}, with schematics shown for both the $ab$ and $ac$-planes in each case and the relevant exchange constants indicated in (a). All the possibilities have Nd moments $\vec{\mu}_{Nd}$~$=$~(1.39(1), 2.25(1), 0)~$\mu_B$. The magnitude of the moment is 2.65(1)~$\mu_B$, which is somewhat lower than the expected value of 3.3~$\mu_B$, but more reasonable than the value of 1.92~$\mu_B$ if one assumes that both Nd sites are long-range ordered. The reduced moment may be a consequence of strong frustration, as it is impossible to satisfy all the $J_1$ interactions within these spin configurations whether they are AFM or FM. 

The presence of the diffuse scattering below $T_N$, as first discussed above, provides further evidence that only one Nd site participates in the long-range order. Furthermore, since the strongest diffuse scattering feature is centered at the same position, $\vec{Q}$~$=$~(1 1/2 1/2), as the most intense, resolution-limited, magnetic Bragg peak, it is likely that the other type of Nd spin chain exhibits quasi-1D, UUDD magnetic order with the spins also confined to the $ab$-plane.  

\begin{figure}
\centering
\scalebox{0.43}{\includegraphics{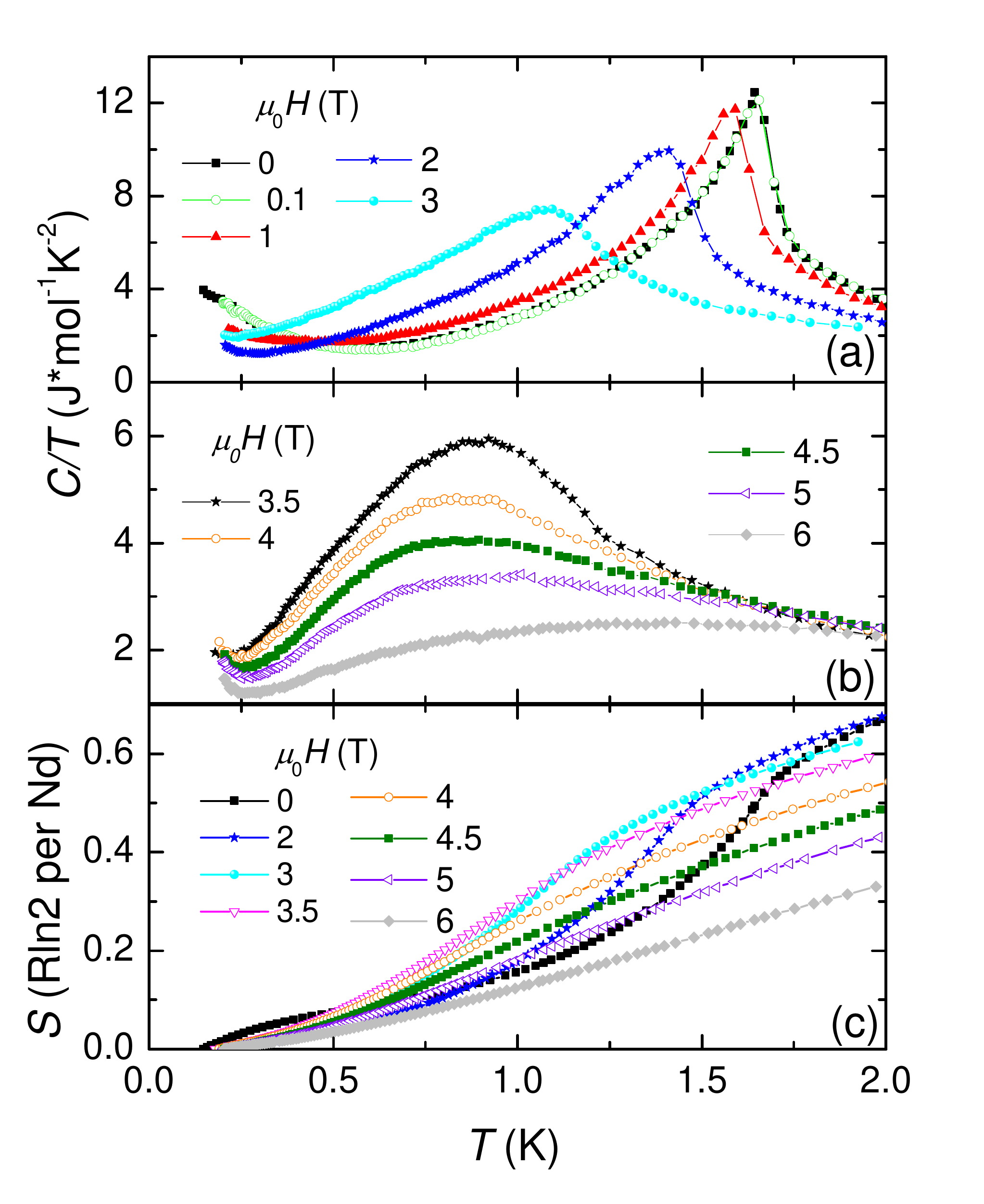}}
\caption{\label{Cp} (a) and (b) show specific heat data of BaNd$_2$O$_4$ as $C/T$ vs. $T$ in magnetic fields between 0~T and 6~T. The sharp $\lambda$ anomaly at 1.7~K in zero field, indicative of a magnetic phase transition, shifts down in temperature in fields up to 4 T. (c) Entropy $S(T)$ curves extracted from the $C_p$ data for specific fields. At the phase transition, less than 5.763~J/mol-K ($Rln2$) of the entropy is recovered, which supports the neutron scattering results that not every Nd atom is involved in the AFM long-range order.}
\end{figure}  

To further characterize BaNd$_2$O$_4$, the specific heat $C_p(T)$ was examined as a function of applied fields $\mu_0 H$. The $\lambda$ anomaly observed in zero field is suppressed with increasing field, as shown in Fig.~\ref{Cp}(a). Figure~\ref{Cp}(b) depicts $C_p$ data for $\mu_0 H$~$>$~3~T where the anomaly at the phase transition shows significant broadening. We interpret this maximum as a Schottky peak for $\mu_0 H$~$\ge$~4~T because it moves to higher temperatures with further increasing fields and it is found where our other measurements indicate that the AFM order is suppressed. Furthermore, we estimate the entropy $S$~$=$~$\int Cp/T\,dT$ from the $C_p$ data, with the results shown in Fig.~\ref{Cp}(c). The jump $\Delta S$ at the phase transition in at all applied fields measured is less than 5.763 J/mol-K ($R ln2$) per individual Nd atom. Assuming a quasi-spin $S$~$=$~1/2 or even higher spin state for Nd, this small entropy gain at the transitions point to incomplete magnetic ordering among the two Nd magnetic sublattices. 

\begin{figure}
\centering
\scalebox{0.43}{\includegraphics{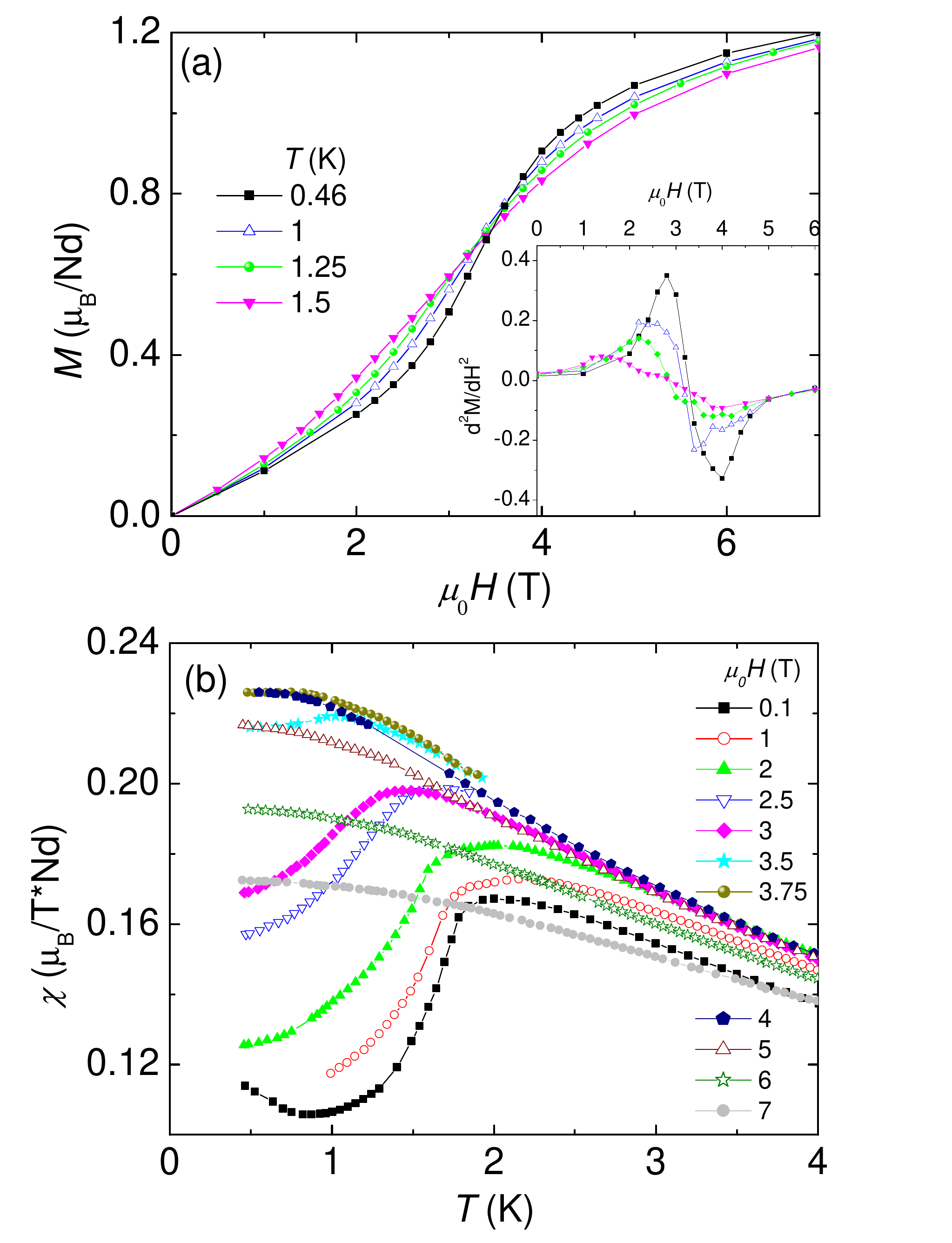}}
\caption{\label{mag} (a) Magnetization vs. field data $M(\mu_0 H)$ for BaNd$_2$O$_4$ at temperatures between 0.46~K and 1.5~K. The inset shows the respective second derivatives $d^2M/d(\mu_0 H)^2$. At 0.46~K, two extrema are apparent in $d^2M/d(\mu_0 H)^2$ that are indicative of the two field-induced phase transitions. These features are not as clear at higher $T$ due to thermal broadening and therefore were not used to map out the phase boundaries in this regime. (b) $T$-dependence of the magnetic susceptibility $\chi$ for magnetic fields between 0 and 7~T. The sharp decrease in the low-field $\chi$ data with decreasing $T$ is a signature of a phase transition from a paramagnetic to an antiferromagnetic state.}
\end{figure}

Magnetization measurements $M(\mu_0 H,T)$ down to 0.46~K complement the neutron scattering and $C_p$ data. Representative $M$ vs. $\mu_0 H$ measurements are shown in Fig.~\ref{mag}(a) with $d^2M/d(\mu_0 H)^2$ vs. $\mu_0 H$ shown in the inset, while $\chi$ (i.e. $M/H$) vs. $T$ is depicted in (b). The $M(\mu_0 H)$ curves in (a) show an S-shape that gets more pronounced with lower temperatures. No full saturation is observed up to 7 T. Instead, the maximum magnetization is only about 1.2~$\mu_B$ per Nd. This is well below the full moment of the free Nd-atom and supports: (i) the results of the neutron scattering experiments and (ii) a smaller than $J$~$=$~9/2 spin state at low temperature, as found in the entropy analysis. Furthermore, two extrema are observed in $d^2M/d(\mu_0 H)^2$ at $\mu_0 H$~$=$~2.75~T and 4~T for $T$~$=$~0.46~K as shown in the inset of Fig.~\ref{mag}(a). Similar features from magnetization data have been used previously to identify field-induced phase transitions in other magnetic systems such as Sr$_3$Cr$_2$O$_8$\cite{09_aczel}, although these extrema in $d^2M/d(\mu_0 H)^2$ only depict the transitions accurately as $T$~$\rightarrow$~0\cite{aczel_thesis}. A similar issue seems to be apparent in the present case, as the extrema are severely broadened with increasing $T$. 

The ordering transitions are also apparent in the $\chi$ vs. $T$ data, defined by a sharp drop in $\chi$ with decreasing $T$. This feature is completely suppressed above 4 T where the magnetic moments are fully polarized, as indicated by a constant $\chi$ behavior for $T$~$\rightarrow$~0. In low magnetic fields such as 0.1~T, a Curie-Weiss contribution to the susceptibility is observed below 0.8~K, possibly arising from a small amount of magnetic impurity or due to the Nd moments of BaNd$_2$O$_4$ that do not participate in the long-range magnetic ordering. However, the moments associated with this signal are fully polarized above 1~T and do not influence our experimental findings.

The most likely spin structures for the intermediate and high-field states correspond to an UUD intrachain configuration and a paramagnetic phase respectively. A field-induced UUD phase has been predicted for classical Ising $J_1$-$J_2$ chain systems with UUDD magnetic structures in zero field\cite{07_heidrich}, and therefore this possibility is in good agreement with conclusions made from the NPD data. We note that the 1/3 magnetization plateau corresponding to the UUD state has already been found in single crystal data of SrR$_2$O$_4$ (R~$=$~Er, Dy, Ho)\cite{12_hayes} when $\mu_0 H$ is applied parallel to the spins within the UUDD chains. In the present case of BaNd$_2$O$_4$, the UUD state should give rise to a 1/3 magnetization plateau when $\mu_0 H$ is applied in the $ab$-plane, but this feature is not expected when $\mu_0 H$~$\parallel$~$\hat{c}$, and therefore the plateau likely gets smeared out in the polycrystalline data. 

\begin{figure}
\centering
\scalebox{0.35}{\includegraphics{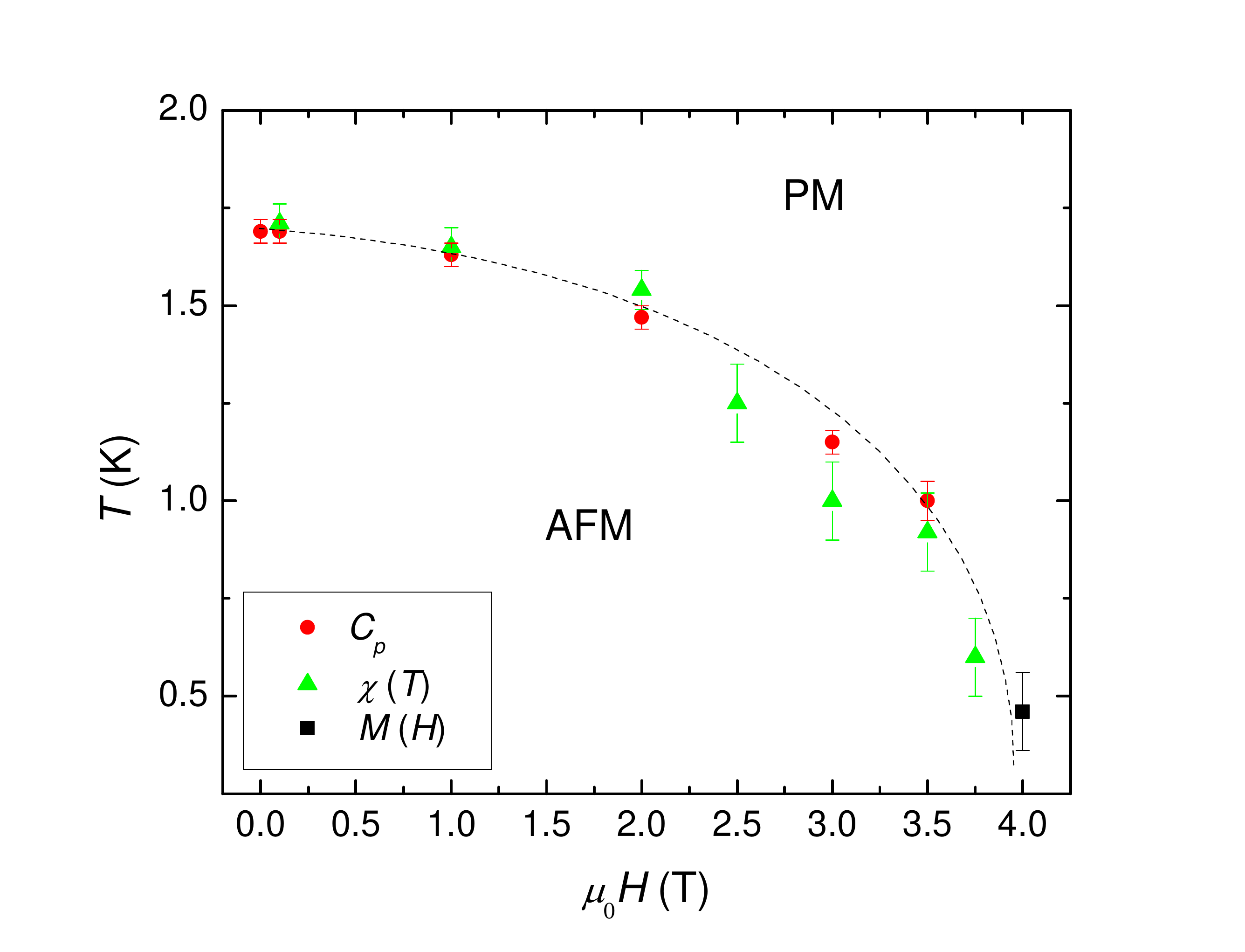}}
\caption{\label{phase_diagram} The phase diagram of polycrystalline BaNd$_2$O$_4$, summarizing the results from $C_p$ (circles), $\chi (T)$ (triangles) and $M(\mu_0 H)$ (squares) measurements. The long-range magnetic order below 1.7~K in zero field is suppressed by 4~T, with the dashed curve serving as a guide to the eye.}
\end{figure}

The combined magnetization and heat capacity data show that a modest magnetic field of only $\mu_0 H$~$=$~4~T is enough to drive the system to a completely paramagnetic state, as summarized in the phase diagram presented in Fig.~\ref{phase_diagram}. The sharp $\lambda$ anomalies observed in the $C_p$ measurements of our polycrystalline sample, combined with the neutron scattering and magnetization results discussed above, imply that the magnetic anisotropy of BaNd$_2$O$_4$ is small but plays an important role in the low temperature magnetic properties. The phase boundary between the low-field UUDD state and the intermediate-field ordered state is not indicated on the phase diagram, as this boundary is expected to be orientation-dependent. The growth of BaNd$_2$O$_4$ single crystals would be highly beneficial for studying the $T$-$\mu_0 H$ phase diagrams along different crystalline directions, so the type of magnetic anisotropy and field-induced phases can be definitively established in this material. 

\section{\label{sec:level4}IV. Summary and Conclusions}

We have investigated polycrystalline samples of the zigzag chain system BaNd$_2$O$_4$ with a combination of magnetization, heat capacity, and neutron diffraction measurements. Neutron diffraction data shows evidence for magnetic Bragg peaks below $T_N$~$=$~1.7~K that can be indexed with a propagation vector of $\vec{k}$~$=$~(0 1/2 1/2). Four candidate magnetic structures are consistent with the data, each arising from only one of the two types of Nd zigzag chains, with the ordered spins lying in the $ab$-plane. All possible magnetic structures are composed of UUDD spin chains, which is exactly the prediction for a classical Ising $J_1$-$J_2$ chain in the large AFM $J_2$ limit. Furthermore, low temperature magnetization and heat capacity measurements as a function of applied field reveal that the order can be completely suppressed in an applied field of only 4 T, with evidence found for an intermediate field-induced state. One possibility is that the intermediate state corresponds to an UUD spin structure, also predicted for the classical Ising $J_1$-$J_2$ chain model. Future inelastic neutron scattering measurements and point charge calculations of the crystal field excitations for the systems in the AR$_2$O$_4$ family would help to clarify the anisotropy of the R atoms and lead to a better understanding of the magnetism in these frustrated magnets. 

\begin{acknowledgments}
This research was supported by the US Department of Energy, Office of Basic Energy Sciences. A.A.A. and V.O.G. were supported by the Scientific User Facilities Division. Work at Los Alamos, J.-Q.Y. and D.M. were supported by the Materials Science and Engineering Division. The neutron experiments were performed at the High Flux Isotope Reactor, which is sponsored by the Scientific User Facilities Division. 
\end{acknowledgments}

\end{document}